\documentclass{jfm}
\usepackage{amsmath,amssymb,bm,array}
\usepackage[usenames,dvipsnames]{color}
\usepackage{graphicx}
\usepackage{natbib}

\newcommand\Rey{\mbox{\textit{Re}}}  % Reynolds number

\title[Periodic orbits and bubbles of chaos]{The role of periodic orbits and bubbles of chaos during the
        transition to turbulence}

\author[S. Altmeyer, A.\ P.\ Willis and B.\ Hof]%
{S.\ Altmeyer$^1$%
  \thanks{Email address for correspondence: sebastian\_altmeyer@t-online.de},\break
A.\ P.\ Willis $^2$ and B.\ Hof$^1$}

\affiliation{$^1$Institute of Science and Technology Austria 
(IST Austria), 3400 Klosterneuburg, Austria\\[\affilskip]
$^2$School of Mathematics and Statistics, University of Sheffield,
Sheffield S3 7RH, UK}

\pubyear{}
\volume{}
\pagerange{}
\date{?; revised ?; accepted ?. - To be entered by editorial office}

\begin{document}

\maketitle 

\begin{abstract}
Starting with turbulence that explores a wide region in phase space, 
we discover several relative periodic orbits (RPOs)
embedded within a subregion of the chaotic turbulent saddle.  We also extract
directly from simulation, several travelling waves (TWs).  
These TWs together
with the RPOs are unstable states and are believed to provide the
skeleton of the chaotic saddle. Earlier studies have shown that such invariant
solutions can help to explain wall bounded shear flows, and a finite subset of them
are expected to dominate the dynamics 
\citep{FaEc2003,PrKe2007,Hofetal2004}. 
The introduction of symmetries is typically necessary to facilitate this
approach.  Applying only the 
shift-reflect symmetry, the geometry is 
less constrained than previous studies in pipe flow.
A `long-period' RPO is identified that is only very weakly repelling.
Turbulent trajectories are found to frequently approach
and frequently shadow this orbit. In addition the orbit
characterises a resulting `bubble' of chaos, itself a saddle,
deep within the turbulent sea \citep{KES2014}.  
We explicitly analyse the merger of the two saddles
and show how it results in
a considerable increase of the total lifetime.
Both {\em exits and entries} to the bubble are observed,
as the stable manifolds of the inner and outer saddles intertwine.
We observe that the typical lifetime
of the turbulence is influenced by switches between the inner and outer
saddles, and is thereby dependent on
whether or not it `shadows' or `visits' the vicinity of the long-period RPO
\citep{Chaosbook}. 
These observations, along with comparisons of flow structures,
show that RPOs play a significant role in structuring the
dynamics of turbulence.
\end{abstract}


%%%%%%%%%%%%%%%%%%%%%%%%%%%%%%%%%%%%%%%%%%%%%%
\section{Introduction}\ref{SEC:intro}
%%%%%%%%%%%%%%%%%%%%%%%%%%%%%%%%%%%%%%%%%%%%%%

Several families of exact coherent structures, i.e.\ invariant sets of solutions,
have been identified in plane Couette flow
\citep{Nag1990,JKSNS2005}, plane Poiseuille flow \citep{Wal2001,Wal2003},
square duct flow \citep{WBN2009,ONWB2010} and the geometry discussed here,
pipe flow \citep{WeKe2004,PrKe2007,FaEc2003}. In general, the state space is
filled with a multitude of unstable invariant solutions that are 
explored by turbulent trajectories. At low flow rates 
the trajectory eventually escapes
from the roller coaster ride through this neighbourhood and ends on
the steady laminar attractor, turbulence is transient here 
and the turbulent neighbourhood corresponds to a {\em chaotic saddle} \citep{KES2014}.
In a circular pipe, the
laminar Hagen--Poiseuille flow is linearly stable at all Reynolds numbers
($\Rey=DU/\nu$, where $D$ is the pipe diameter, $U$ the mean axial velocity
and $\nu$ the kinematic viscosity of the fluid). Several
families of three-dimensional travelling wave (TW) solutions have been
discovered \citep{FaEc2003,WeKe2004}, which represent the `simplest' invariant
solutions in pipe flow satisfying
\begin{eqnarray}
  \label{EQ:TW}
  {\bm u}(r,\theta,z,t) = {\bm u}(r,\theta,z-ct),
\end{eqnarray}
where $(r,\theta,z)$ are the usual cylindrical coordinates, ${\bm u} =
(u,v,w)$ are the corresponding velocity components, $t$ the time and $c$ the
wave speed.  The TW solutions originate from saddle-node bifurcations at a
finite value of the Reynolds number.  Mellibovsky and Eckhardt
\citep{MeEc2011} provide a fundamental study of TWs' origins and their subsequent
varied bifurcations. Periodic solutions on the other hand bifurcate classically
in a Hopf bifurcation out of TWs. Orbits are important as they capture the
dynamics. In a recent study of transition in Couette flow, Kreilos
and Eckhardt \citep{KeEc2012}, found that such orbits undergo a 
transition sequence to chaos. They
followed the bifurcation of exact coherent states that undergo period-doubling
cascades and end with a crisis bifurcation.  
Due to the strong advection of structures in pipe flow, 
only {\em`relative'} periodic orbits (RPOs) are observed.
These orbits include a streamwise translation with mean phase speed 
$\overline{c}$.
Of these, only few have been discovered so far \citep{DPK2008a,WCA2013,AMRH2013}.  
They are expected to capture the natural measure of turbulent flow
\citep{CiGi2010,WCA2013}. Such RPOs satisfy
\begin{eqnarray}
  \label{EQ:RPO}
  {\bm u}(r,\theta,z,t) = {\bm u}(r,\theta,z-\overline{c}t,t+T),
\end{eqnarray}
such that the motion appears $T$-periodic in a frame co-moving at speed
$\overline{c}$.  The value $\overline{c}$ is different for
each RPO.  Similar structures were also reported in plane Couette flow
\citep{KaKi2001,KaSa2005,Vis2007}.
To date, the number of RPOs
discovered in pipe flow is few
and all searches were limited to subspaces with rotational symmetries \citep{MeEc2011,WCA2013}.
It is worth mentioning,
however, that any solution found in a subspace are necessarily also
solutions of the full space and hence represent physically consistent
flow states. All RPOs
discovered were embedded in regions of `lower' energy (below turbulent levels) but it was 
speculated that RPOs at higher energy levels exist that underpin the 
dynamics of turbulence.  

All known invariant solutions other than the laminar flow
are unstable at the Reynolds
numbers for which turbulence is observed, but the dimensions of their
unstable manifolds in phase space is typically low
\citep{Kaw2005,KeTu2007,Wal2001,Vis2009}. Hence it is expected
that they can be approached closely along their stable manifolds.
The least unstable orbits are expected to be the most representative,
and are the most likely to be extracted from simulation.

In this paper we isolate 
and link significant features of turbulent dynamics
to several representative RPOs,
including one with a much longer period time than
previously computed orbits.  We present results of invariant
solutions for a `minimal' practical set of symmetry, imposing only
shift-reflect and no rotational symmetry.
When the long-period RPO emerges
it forms a localised bubble of chaos within the rest of turbulence.
Shadowing, or `visits' to this invariant solution significantly 
increase the turbulent lifetime.
The resulting lifetime of a trajectory is dependent on the rate of 
switching, that is, we observe `entries' and `exits' from the bubble.
In addition, the
long-period RPO captures much of the qualitative features of
turbulence, including streak break-up and repetitions in chaotic trajectories.

%%%%%%%%%%%%%%%%%%%%%%%%%%%%%%%%%%%%%%%%%%%%%%%%%%%%%%%%%%%%%%%%%%%%%%
\section{Numerics and Formulation}\label{SUBSEC:munerics}
%%%%%%%%%%%%%%%%%%%%%%%%%%%%%%%%%%%%%%%%%%%%%%%%%%%%%%%%%%%%%%%%%%%%%%

To find invariant states we first apply the symmetry reduction {\em method
of slices} \citep{WCA2013,BCDS2014} to pipe flow to obtain a quotient of the
streamwise translation symmetry of turbulent flow states.  Within the
symmetry-reduced state space, all TWs reduce to equilibria and all RPOs
reduce to periodic orbits.  The method bypasses the difficulties of
$\bar{c}$ being different for each RPO in an automatic manner, 
and permits much
simpler identification and extraction of TWs and RPOs directly from
turbulent chaotic trajectories. 
For simulations we use a hybrid spectral finite-difference code
\citep{WiKe2009} with 64 non-equispaced finite difference axial points, Fourier
expansions evaluated on 48 azimuthal points and on 24 points per unit radius in
the radial direction (idealised).  The shift-reflect symmetry, carried by
almost all known TWs, is applied,
\begin{eqnarray}
  \label{EQ:reflectional_sym}
  {\bm S}:(u,v,w)(r,\theta,z,t) = (u,-v,w)(r,-\theta,z,t).
\end{eqnarray}
For convergence and continuation of solutions we used a
Newton-Krylov-hookstep algorithm \citep{Vis2007} with minor enhancements in
adjustments to the norm \citep{WCA2013}.  The relative
residual of the RPOs is at least approximately $10^{-7}$ for
the longest orbit, and is considerably less ($<10^{-8}$) for the others.  While
we have not imposed further symmetries, some of the observed TW solutions are
highly-symmetric N1 \citep{PDK2009}, which also satisfy shift-and-rotate
symmetry
\begin{eqnarray}
  \label{EQ:shift_rotate_sym}
  {\bm \Omega}:(u,v,w)(r,\theta,z,t) = (u,v,w)(r,\theta+\pi,z-\pi/\alpha,t),
\end{eqnarray}
where $2\pi/\alpha$ ($\alpha=1.25$) is the wavelength.  
(See \citep{WCA2013} Appendix regarding relationships between 
symmetries in pipe flow.)  For the present work we
fixed $Re=2300$ and have chosen a domain of length $5D$, which is a
compromise between the reduced computational expense of a 
smaller domain and the need for the pipe to be sufficiently long to accommodate
turbulent dynamics.  We note that it has recently been shown that the key
features of localised TWs are easily encompassed within a domain of similar
length in pipe flow \citep{CWK2014}.  For these parameters, turbulence is
found to be transient with characteristic life-time $\bar{t}\approx
10^3D/U$. % \citep{WCA2013}.

%%%%%%%%%%%%%%%%%%%%%%%%%%%%%%%%%%%%%%%%%%%%%%%%%%%%%%%%%%%%%%%%%%%%%%
\section{Relative Periodic Orbits}\label{SEC:results}
%%%%%%%%%%%%%%%%%%%%%%%%%%%%%%%%%%%%%%%%%%%%%%%%%%%%%%%%%%%%%%%%%%%%%%

\begin{figure}
\begin{center}  
\begin{tabular}{cc}
\raisebox{0.4\linewidth}{(a)}&
\includegraphics[width=0.6\linewidth]{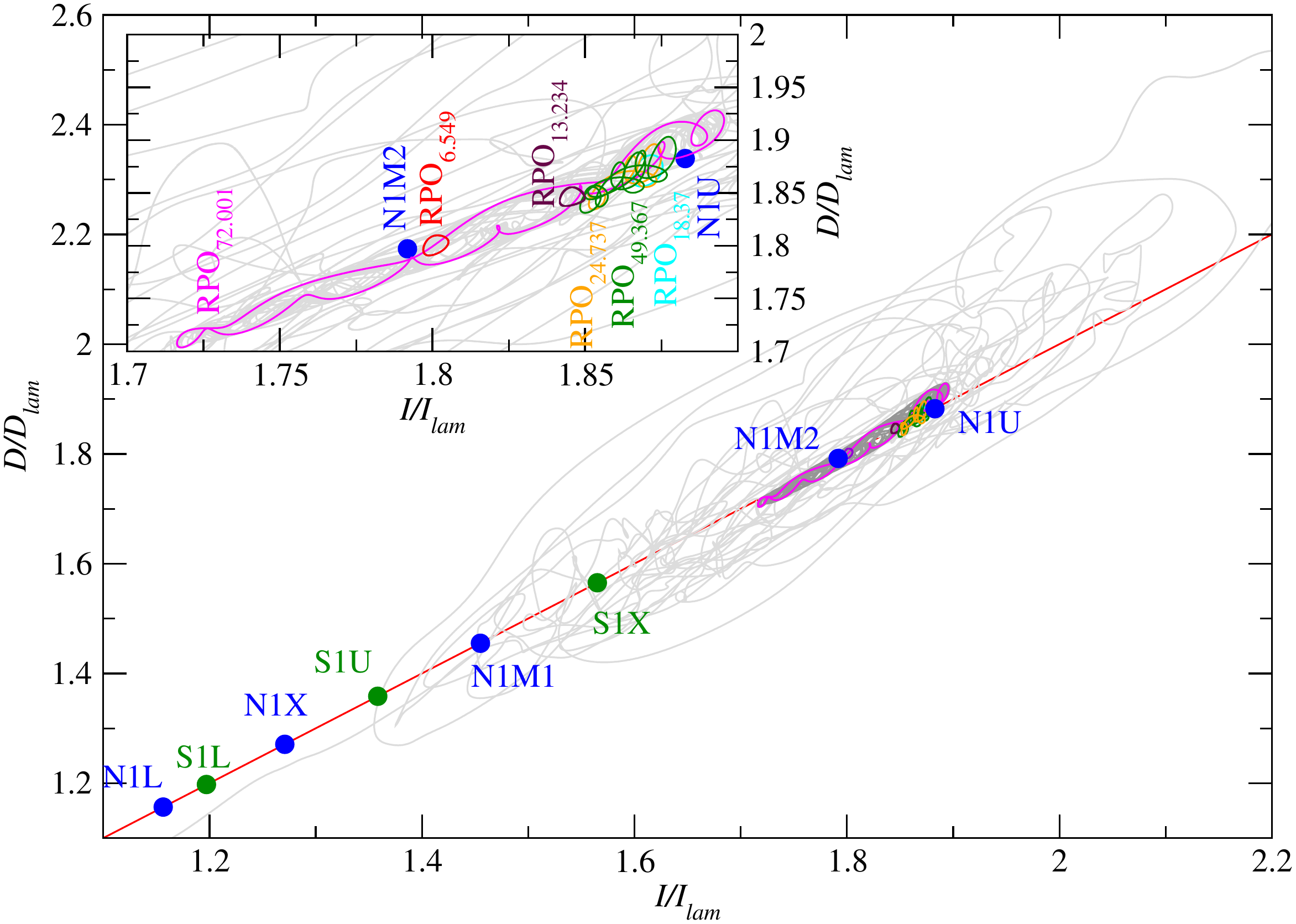}\\
\raisebox{0.4\linewidth}{(b)}&
\includegraphics[width=0.6\linewidth]{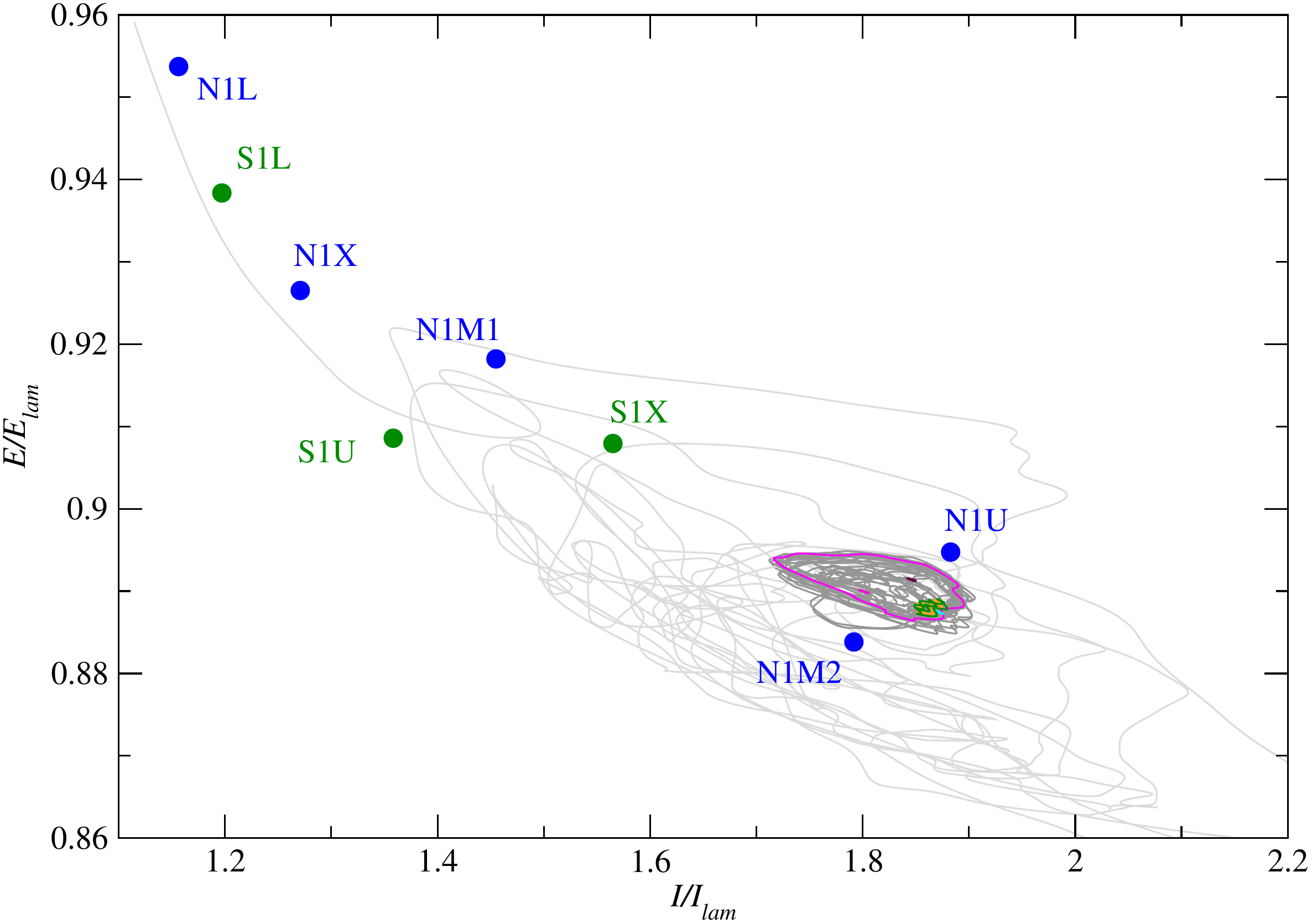}\\
\raisebox{0.4\linewidth}{(c)}&
\includegraphics[width=0.6\linewidth]{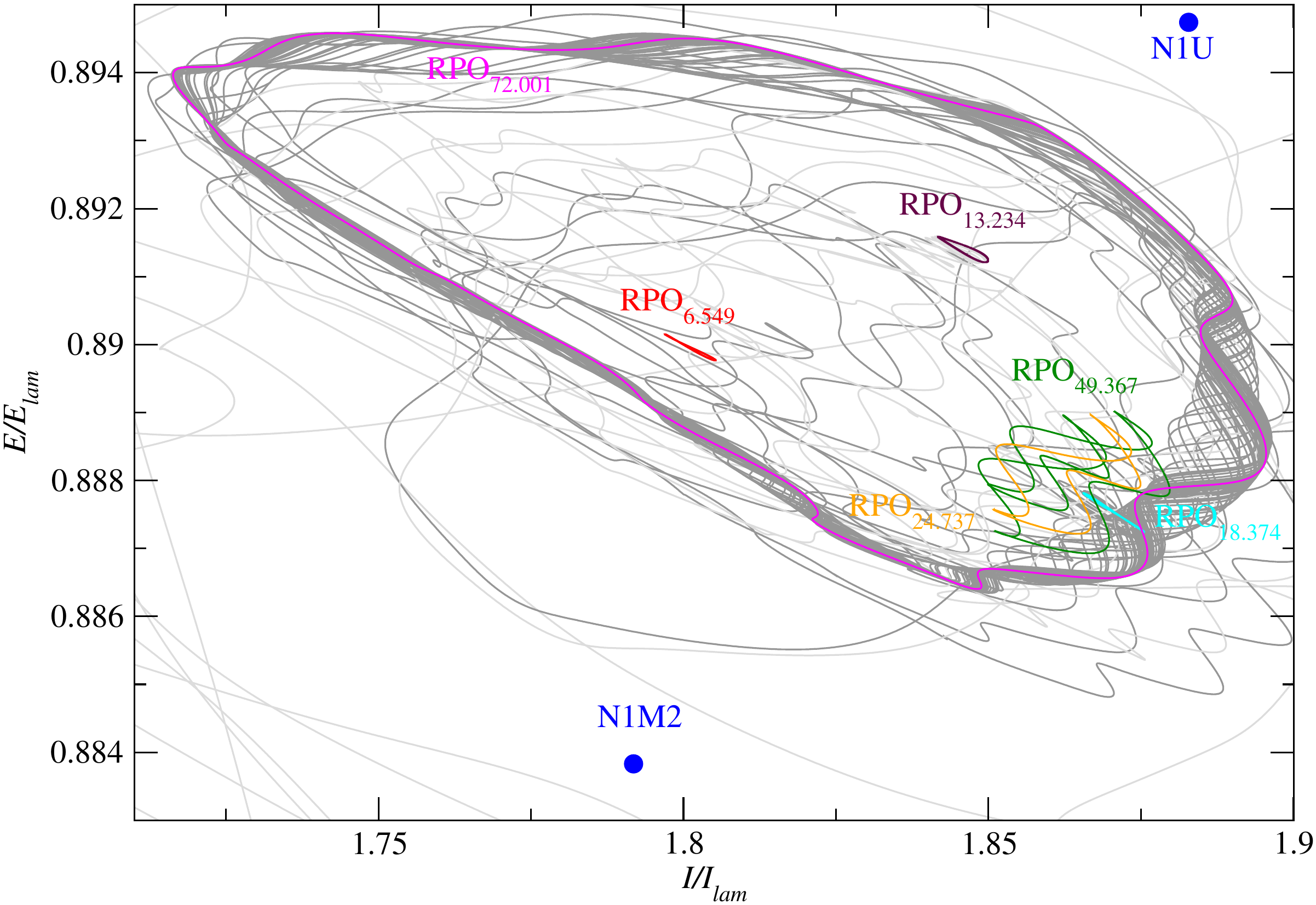}
\end{tabular}
\end{center}
\caption{
         Rate of energy input from the background pressure
         gradient $I$ (external power
         to maintain constant flux) versus $(a)$ the dissipation rate
         $D$ and $(b)$ the
         energy $E$ for the invariant
         solutions of TWs and RPOs, together with two turbulent orbits. 
         (All quantities are normalised by their laminar counterparts.)
         Light gray presents a typical turbulent orbit while dark gray
         indicates orbits that shadow RPO$_{72.001}$ and enter and exit the bubble.
         $(c)$ and inset in $(a)$ show an expanded view of the region near
         N1M2 and N1U where all here discovered RPOs are embedded. Note, due
         to visibility a full relaminarisation trajectory is only given for
         one simulation.
        }
\label{fig:D-E-I}
\end{figure}

The classical physical quantities, input energy $I=V^{-1}\oint dS[{\bm n}\cdot{\bm u}]p$,
dissipation $D=||{\bf\nabla}\times{\bm u}||_2^2/Re$ and
the kinetic energy $E=||{\bm u}||_2^2/2$, 
where $||\cdot||$ corresponds to a root-mean-square value, 
are often used for phase space representation.
Figure \ref{fig:D-E-I} presents projections of several
travelling waves and the relative periodic
orbits.  For notation we follow the works
\citep{KeTu2007,PDK2009}. Highly-symmetric TWs with both
shift-reflect and shift-rotate symmetry ($S,\Omega$) are named `N1',
and states with only shift-reflect symmetry ($S$) are called `S1'.
We identify different RPOs
with subscript corresponding to their period.
(Note that
these numbers refer to $Re=2300$ and change with $\Rey$ \citep{WCA2013}.) Due
to energy balance, all TWs, time-averages of RPOs and a sufficiently long time-average of a turbulent flow
must lie on the diagonal $D=I$ (see Fig.\ \ref{fig:D-E-I}$(a)$ ). 
While the lower branch travelling waves N1L and S1L (see Fig.  \ref{fig:D-E-I}$(a)$)
are far from the turbulent flow, the TWs, N1M2 and N1U
appear to be in the core of the turbulent region (Although the classes `N'
and `S' are well known
most of `1-fold' solutions observed here are new.)
Notably, all RPOs extracted from turbulent trajectories also lie within this
region (in particular near the upper branch TW of N1U) while all short RPOs
are enclosed by the long-period RPO$_{72.001}$ in the
energy plot (see Fig.\ \ref{fig:D-E-I}$(c)$).  
Extraction of these orbits from simulation implies that they
are among the least unstable, and are therefore expected to be important and potentially
representative of turbulent flow.  In particular the long-period 
RPO$_{72.001}$ is found to be dynamically relevant.
Trajectories that happen to approach the orbit tend to stay close to it and
encircle it (see dark regions in Fig.\ \ref{fig:D-E-I}) for extended periods.
This results in an overall increase in lifetimes.
All RPOs discussed here lie in the core region, associated with higher dissipation,
and underline their potential role as organising centers of the turbulent dynamics. In particular our RPOs are located
near the solutions of upper branch TWs,
rather than the lower branch TWs, the latter of which are more closely linked to transition.

\begin{figure}
\begin{center}  
\begin{tabular}{cc}
\raisebox{0.4\linewidth}{(a)}&
\includegraphics[width=0.6\linewidth]{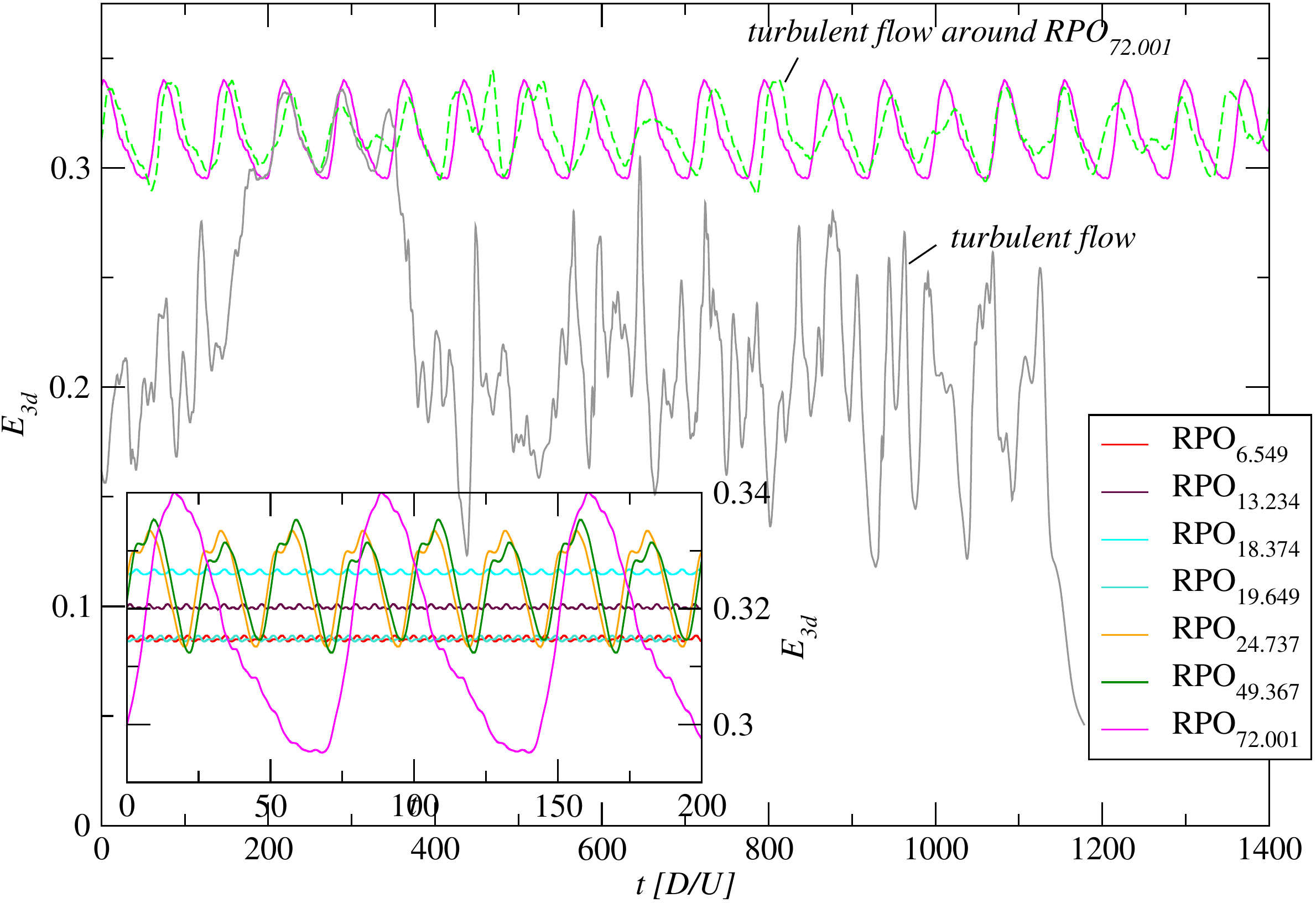}\\
\raisebox{0.4\linewidth}{(b)}&
\includegraphics[width=0.6\linewidth]{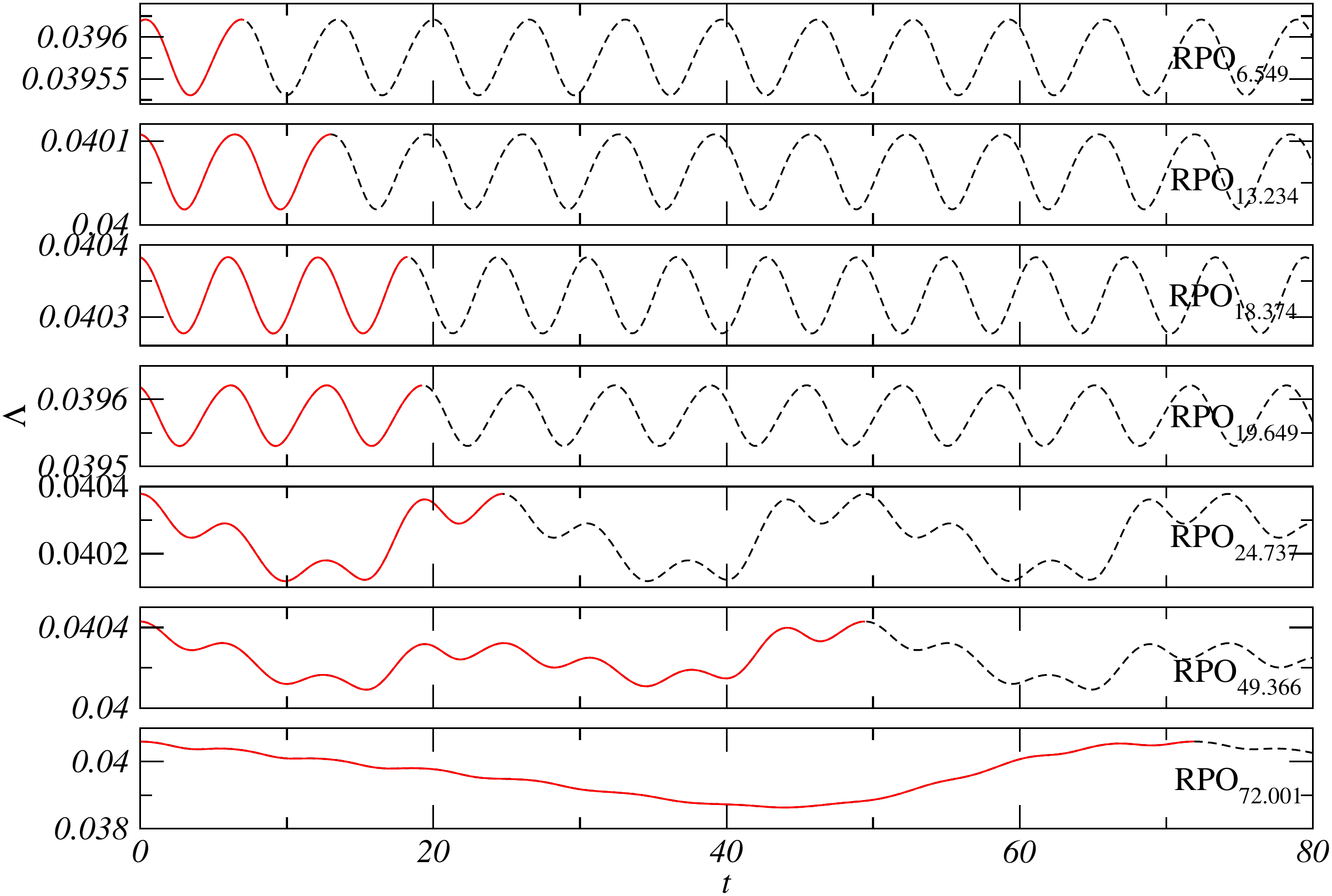}
\end{tabular}
\end{center}
\caption{
         Dynamics of pipe flow for several RPOs and two
         turbulent runs.  For the latter, one (green (dark gray) dashed) hangs around
         the long-period RPO$_{72.001}$ (stuck in inner saddle) while the other shows a typical visit of
         the RPO$_{72.001}$ orbit (here about two periods; typically about two to five
         periods) before it relaminarizes.  $(a)$ Energy 
         $E_{3d}$
         versus time $t[D/U]$ for each RPO.  The inset shows the periodic
         oscillation of all RPOs discussed in this letter.  $(b)$ Friction
         of RPOs as indicated.  One cycle of each period is shown red (light gray)
         solid, with continuation black dashed. 
        }  
\label{fig:E-t_RPOs}
\end{figure}

Figure \ref{fig:E-t_RPOs} illustrates the close correspondence between
the dynamics of turbulent runs and those of RPO$_{72.001}$.
Monitoring the energy $E_{3d}=||{\bm u}-{\bar{\bm u}}||_2^2/2$ (top panel),
where ${\bar{\bm u}}$ is the axially averaged flow, and
friction $\Lambda=:2DG/\rho U^2$ (where $G$ is the mean pressure gradient
along the pipe and the density)
(bottom panel) against time $t$ we show and example 
trajectory (green, dashed) that stays around
the long-period RPO$_{72.001}$ for a very long time, 
and another (gray) that shows
a more typical `visit' to RPO$_{72.001}$, here for approximately two periods.
The group of periodic orbits is shown in the inset of Fig. 
\ref{fig:E-t_RPOs}$(a)$ and $(b)$. The frequency of the shortest orbit
RPO$_{6.549}$ appears as a fast frequency modulation of the longer period orbits.
Closeness of the two orbits
RPO$_{24.737}$ and RPO$_{49.367}$ is clear in both energy and friction (Fig. 
\ref{fig:E-t_RPOs}) and in the phase-space plot (Fig.\ \ref{fig:D-E-I}).
Compared to RPOs found by numerical continuation from bifurcations, the
range of energy variation $\Delta E_{3d}$ is large \citep{WCA2013}, and
the long period orbit RPO$_{72.001}$ explores a wide area deep in the turbulent sea 
(Fig.\ \ref{fig:D-E-I}$(b)$).  The orbits with the three next longer periods only
show a week longer period modulation on top of the fast frequency. The three orbits
with largest periods are modulated much more strongly. Equally they explore 
a wider range covered by the turbulent trajectories.
Construction of orbits from shorter ones, and that longer orbits
show rather complex structural changes and
explore a wider range, supports the idea that RPOs function as building
blocks of the turbulent motions.  
While RPO$_{24.737}$ and RPO$_{49.366}$ show several
local maxima over one period, the longer RPO$_{72.001}$
is only very weakly modulated (one single peak, slowly varying over one full period).
Independent of their period time, all these RPOs show the same underlying
short-time modulation with approximately the period of the shortest one RPO$_{6.549}$ (Fig.\ \ref{fig:E-t_RPOs}$(b)$).
Here it seems likely that all RPOs discussed may be members of the same family.

\begin{figure}
\begin{center}  
\includegraphics[width=0.6\linewidth]{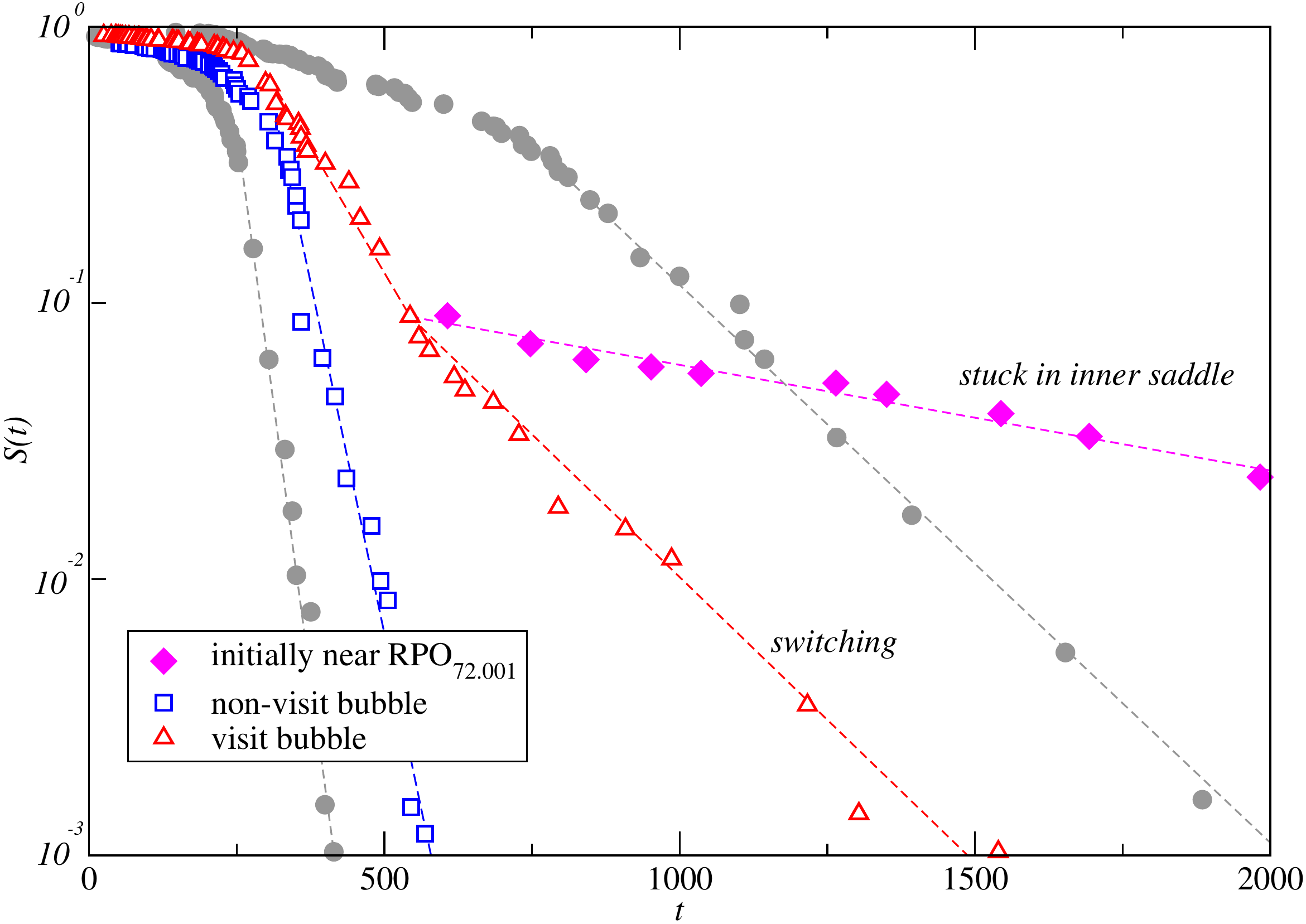}
\end{center}
\caption{
         $(a)$ Relaminarisation times of turbulent flows at $Re=2300$
         as a function of initial conditions (times in $D/U$) separated if its
         trajectories visit ($\textcolor{red}{\vartriangle}$) and non-visit
         ($\textcolor{blue}{\square}$) the vicinity of RPO$_{72.001}$ orbit
         (horizontal dashed lines present averaged lifetimes) and initials near that
         orbit ($\textcolor{magenta}{\blacklozenge}$) which get stuck in the
         inner saddle. $(b)$ Corresponding survivor function
         $S(t)=\exp[(t-t_0)/\tau_\text{true}]$, with
         $\tau_\text{true}=1/r[\sum_{i=1}^r t_i+(n-r)t_r]$ (lifetime sample
         of size $n$ with truncation after $r$ decays) \citep{AWH2010}.
         The attractor of RPO$_{72.001}$ cause an offset to
         `nearby' initial conditions that decay after
         longer times. `Switching' refers to changes between inner and outer
         saddle. For comparison, results for $Re=2250$ and $Re=2450$ are also 
         shown, which are below and above the range existing where RPO$_{72.001}$
         exists ($2292 \lesssim Re\,(\text{RPO}_{72.001}) \lesssim 2423$).          
        }  
\label{fig:surv_function}
\end{figure}

To quantify the effect of the bubble we calculated 
relaminarisation times of turbulent flows for about 300 
initial conditions (ICs). These were separated in terms of those
that `visit' transiently the inner saddle or `bubble' near
RPO$_{72.001}$, and other trajectories which `non-visit'.
The `bubble' near RPO$_{72.001}$ is a saddle within the outer turbulent saddle 
that is sufficiently large to cover 
the dark gray region in Fig. \ref{fig:D-E-I}.
Large fluctuations in the data indicate that the ICs have
uncorrelated lifetimes, but the lifetimes are longer for turbulent flows that
visit the bubble. Based on these ICs we calculated
the corresponding survivor functions $S(t)$ (Fig. 
\ref{fig:surv_function}). The exponential distribution
$S(t)=\exp[(t-t_0)/\tau_\text{true}]$ estimate the characteristic turbulent lifetimes
with the sample mean, which is the maximum likelihood estimator (MLE) of the
parameter $\tau_\text{true}$. ICs started in the very close neighbourhood of the orbit 
RPO$_{72.001}$ spend very long
time within this subregion
(labelled `stuck in inner saddle'), but are otherwise uncorrelated.
While their distribution follows that of a memoryless process,
their long period results in distribution that is almost
the superposition of an exponential and a constant, similar to the observation
in \citep{KES2014}, suggesting that they are associated
with a further saddle within the inner saddle (or another `bubble' within the bubble).
More typically, all other ICs lead to trajectories that visit the bubble near
RPO$_{72.001}$ for two to five periods, and the extended lifetime is
dependent on the rate of `switching', or `entries' and `exits'
from the bubble.

Comparing before and after the range where
the long period orbit RPO$_{72.001}$ exists 
($2292 \lesssim Re\,(\text{RPO}_{72.001}) \lesssim 2423$) in
Fig.\ \ref{fig:surv_function} (gray circles),
it is observed that at slightly higher $Re=2450$ the flow
has inherited the longer lifetime associated with the bubble
(compare with slope for the `switching' case).  
The bubble has fully merged with the outer saddle
so that the spuriously long lifetimes
associated with the possible saddle within in the inner saddle,
no longer feature.

Our simulations highlight that visits, or `switches',
to the `bubble' near RPO$_{72.001}$ are
typically of around two to five period times, i.e.\ `shadowing' this orbit
two to five times, see Fig.\ \ref{fig:E-t_RPOs}(a).
Thus visiting the bubble strongly affects the
turbulent lifetimes suggesting that the RPOs that structure this region
play a key role in the evolution of the turbulent flow.
Notably also,
perturbations of the shorter RPOs results in a turbulent flow that also
remain in this region and around RPO$_{72.001}$ for substantial times.
This strongly suggests further `shadowing' of the orbits, 
and that longer orbits may be constructed in terms of shorter ones.

\begin{figure}
\begin{center}
  \begin{tabular}{cc}
   \raisebox{0.3\linewidth}{$(1)$}&   
   \includegraphics[width=0.45\linewidth]{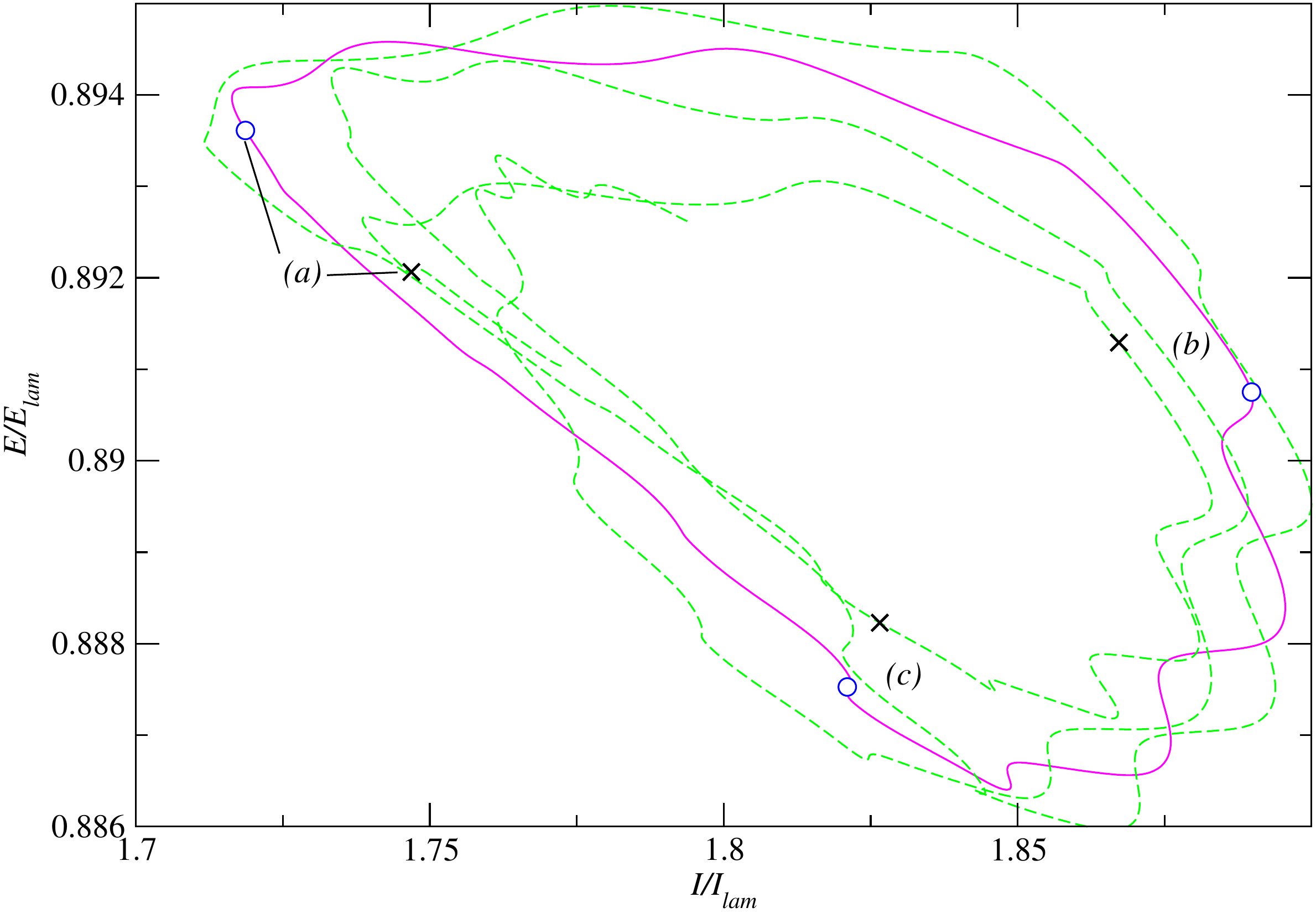}\\
   \raisebox{0.3\linewidth}{$(2)$}&   
   \includegraphics[width=0.45\linewidth]{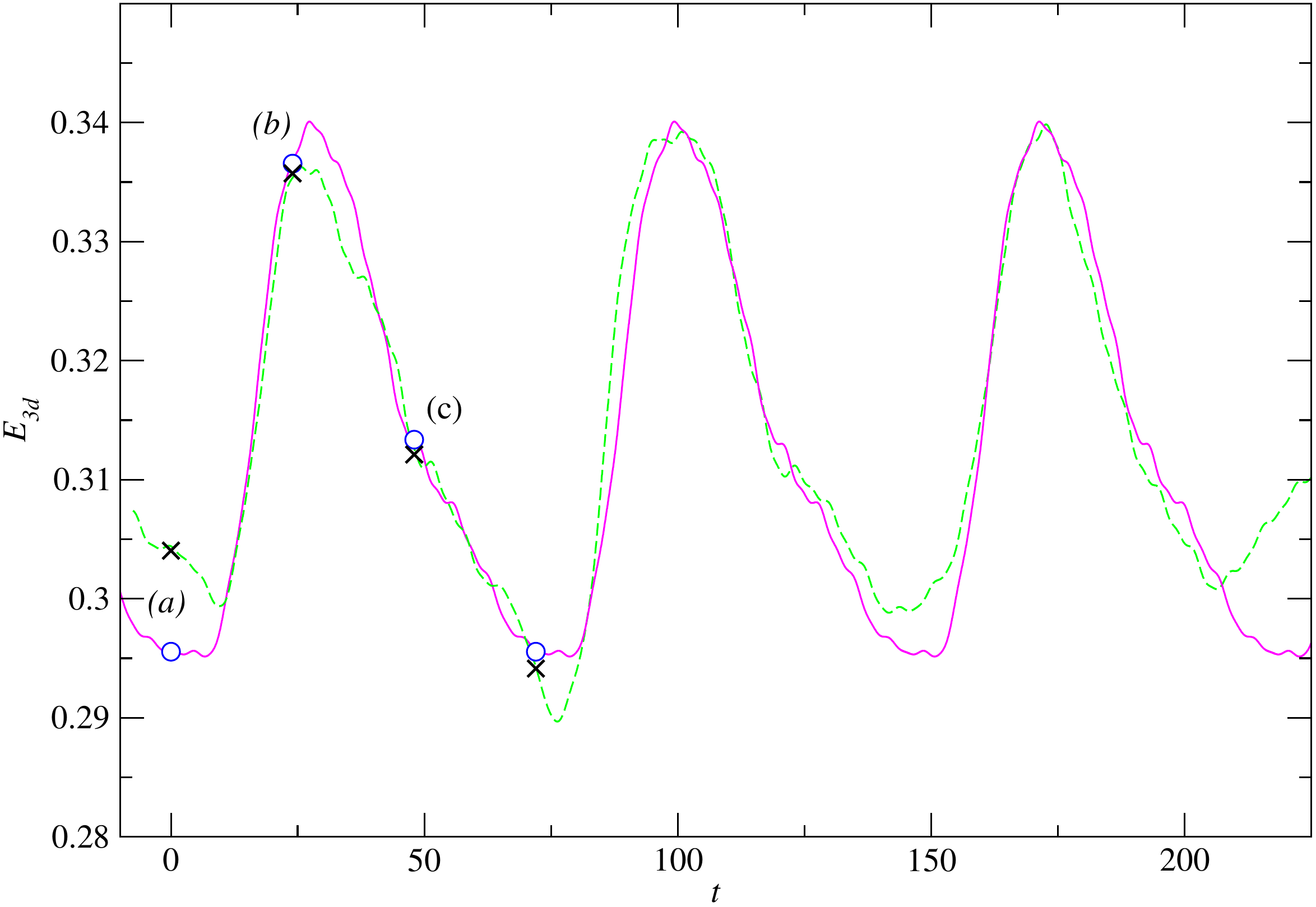}
  \end{tabular}   
\end{center}   
  \begin{tabular}{ccc} 
   & RPO$_{72.001}$ & turbulent flow \\
   \raisebox{0.18\linewidth}{$(a)$}&
   \includegraphics[width=0.45\linewidth]{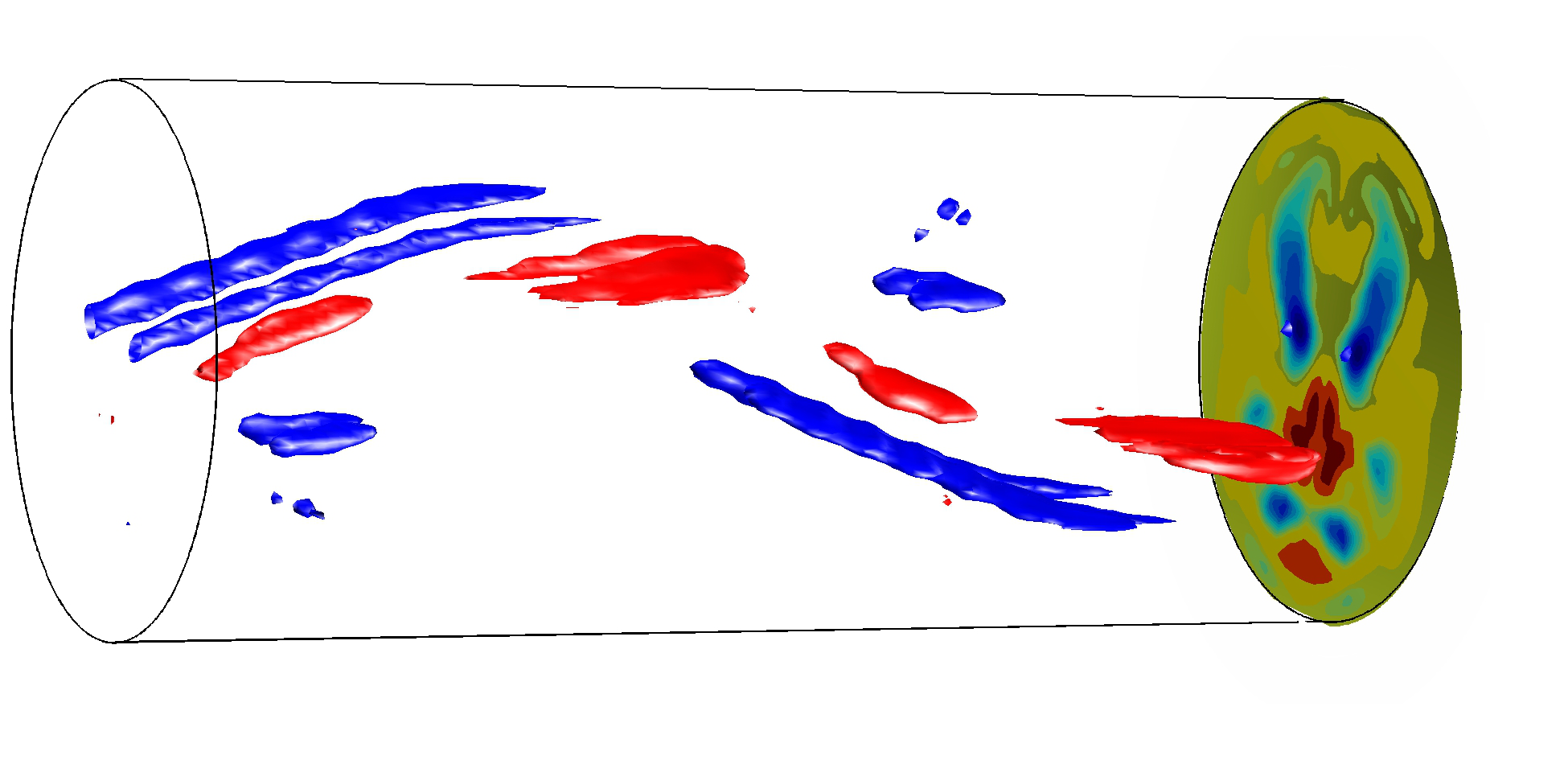}&
   \includegraphics[width=0.45\linewidth]{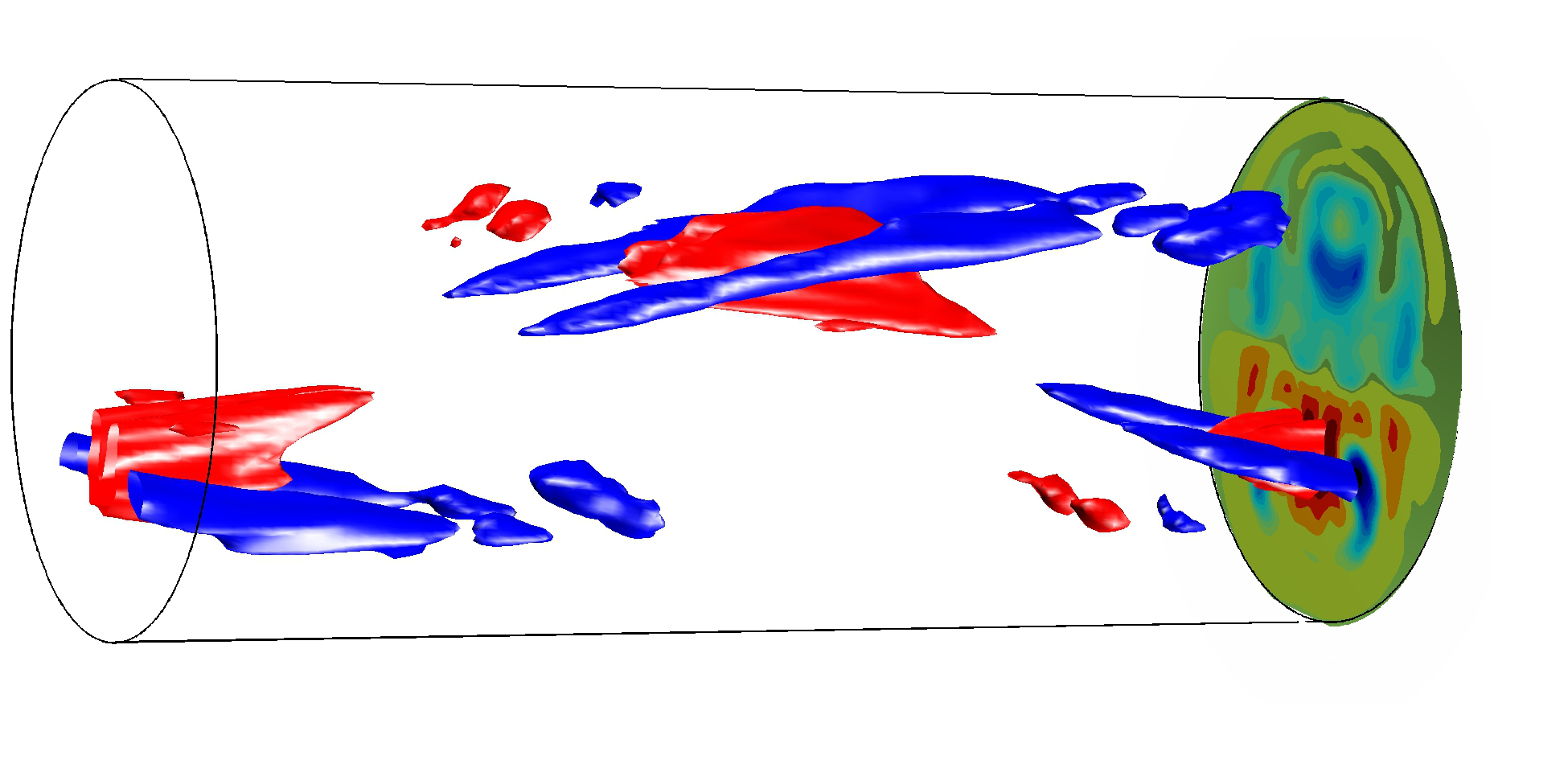}\\
   \raisebox{0.18\linewidth}{$(b)$}&   
   \includegraphics[width=0.45\linewidth]{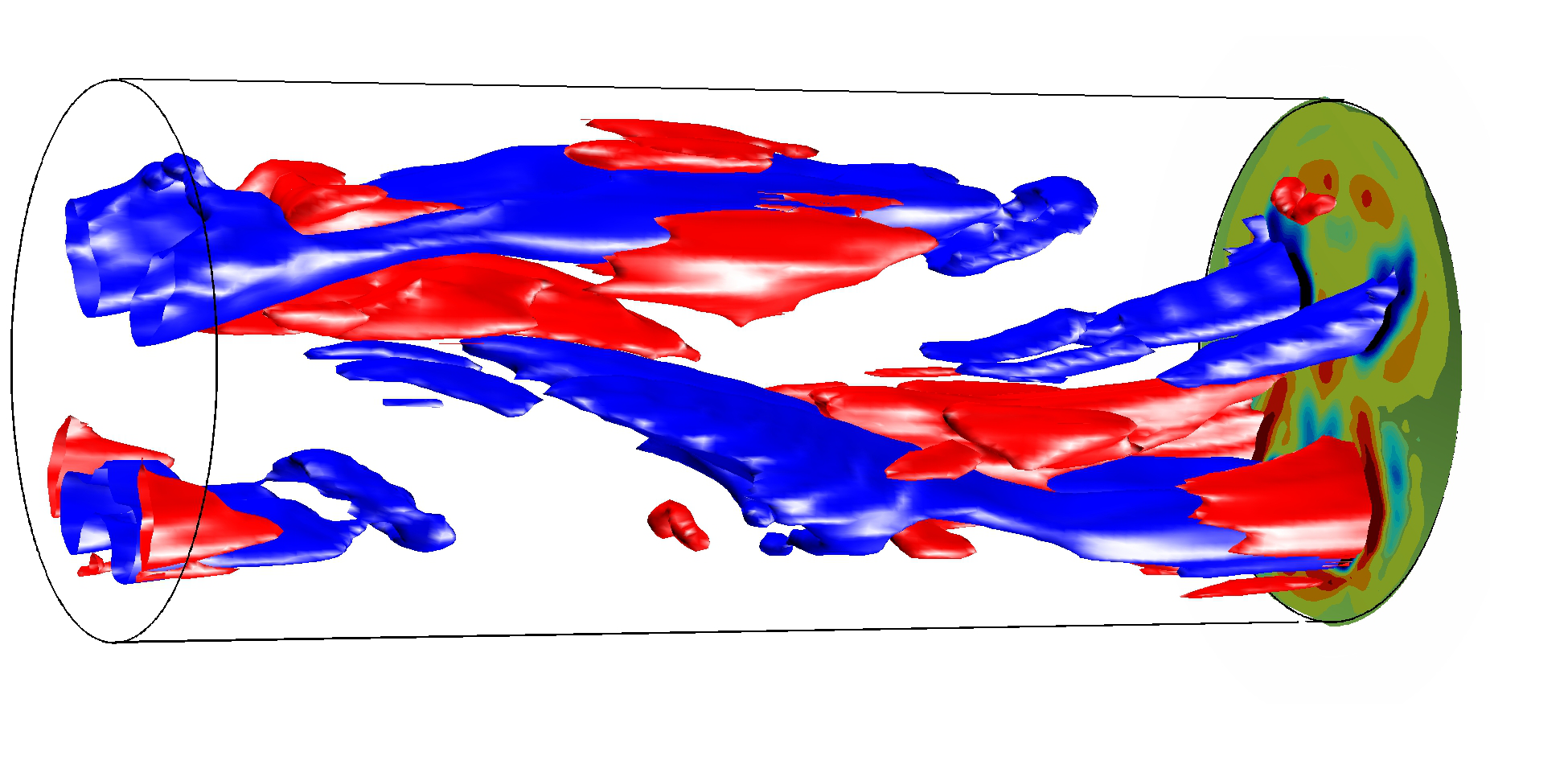}&
   \includegraphics[width=0.45\linewidth]{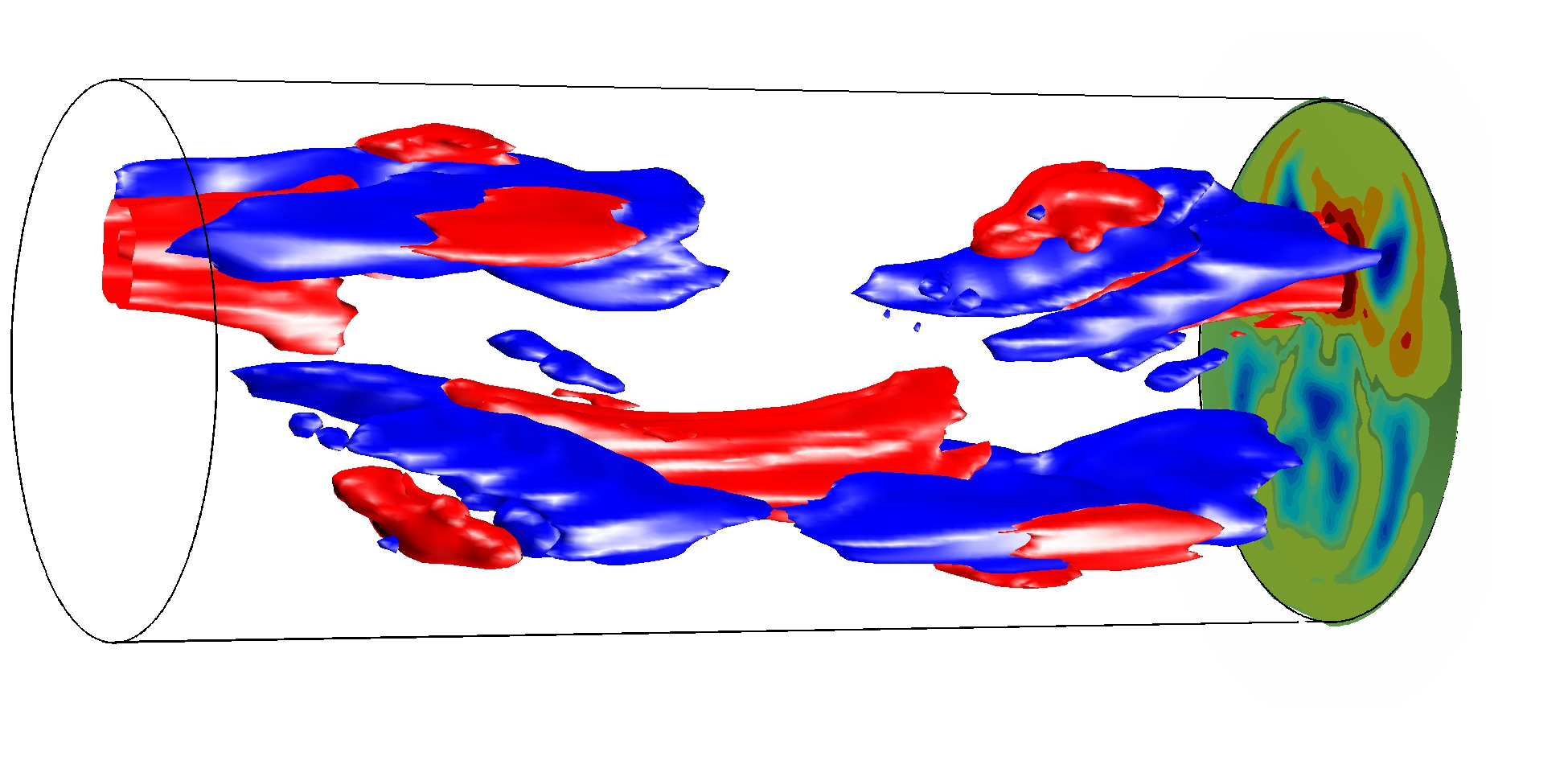}\\
   \raisebox{0.18\linewidth}{$(c)$}&   
   \includegraphics[width=0.45\linewidth]{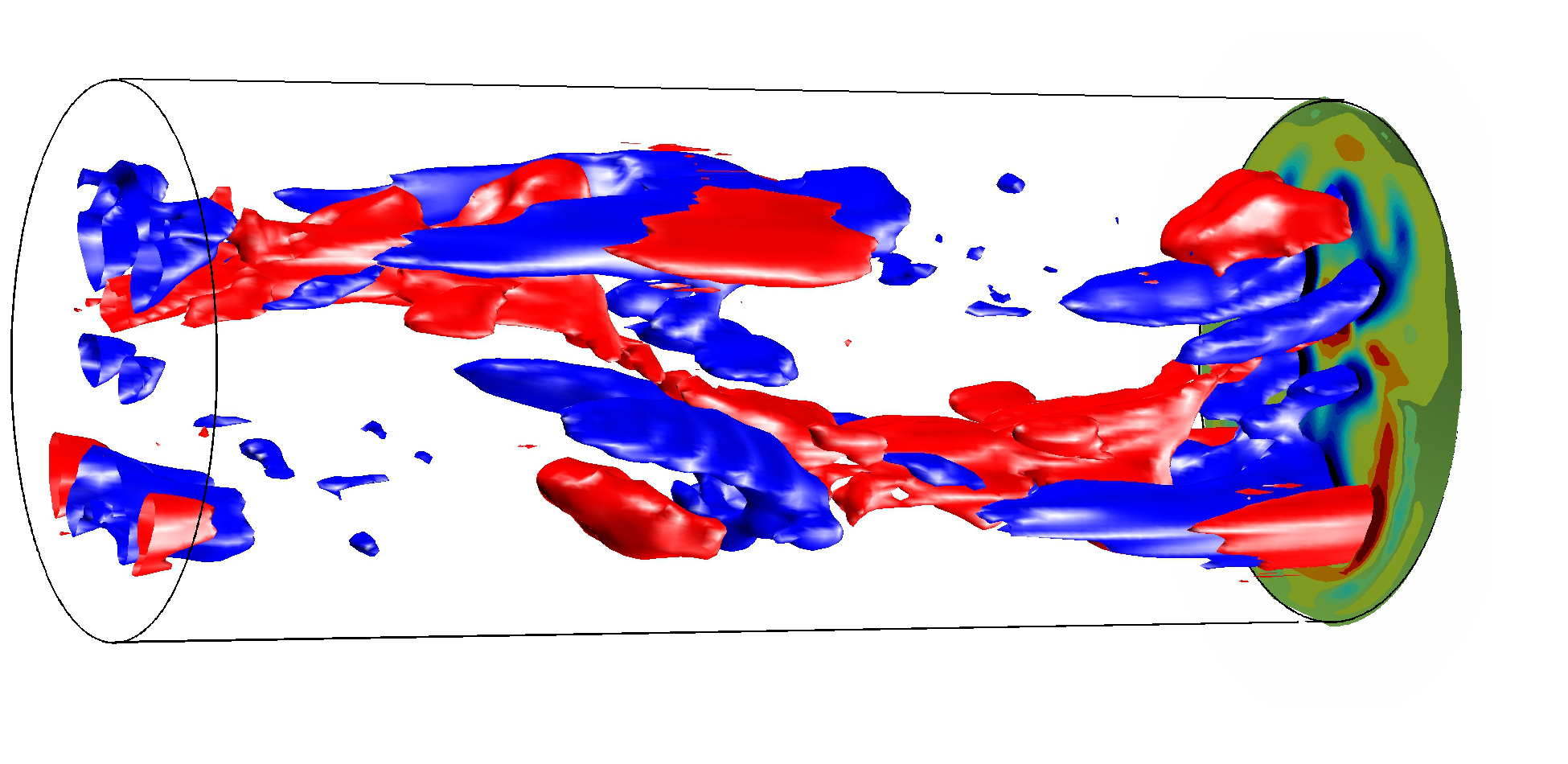}&
   \includegraphics[width=0.45\linewidth]{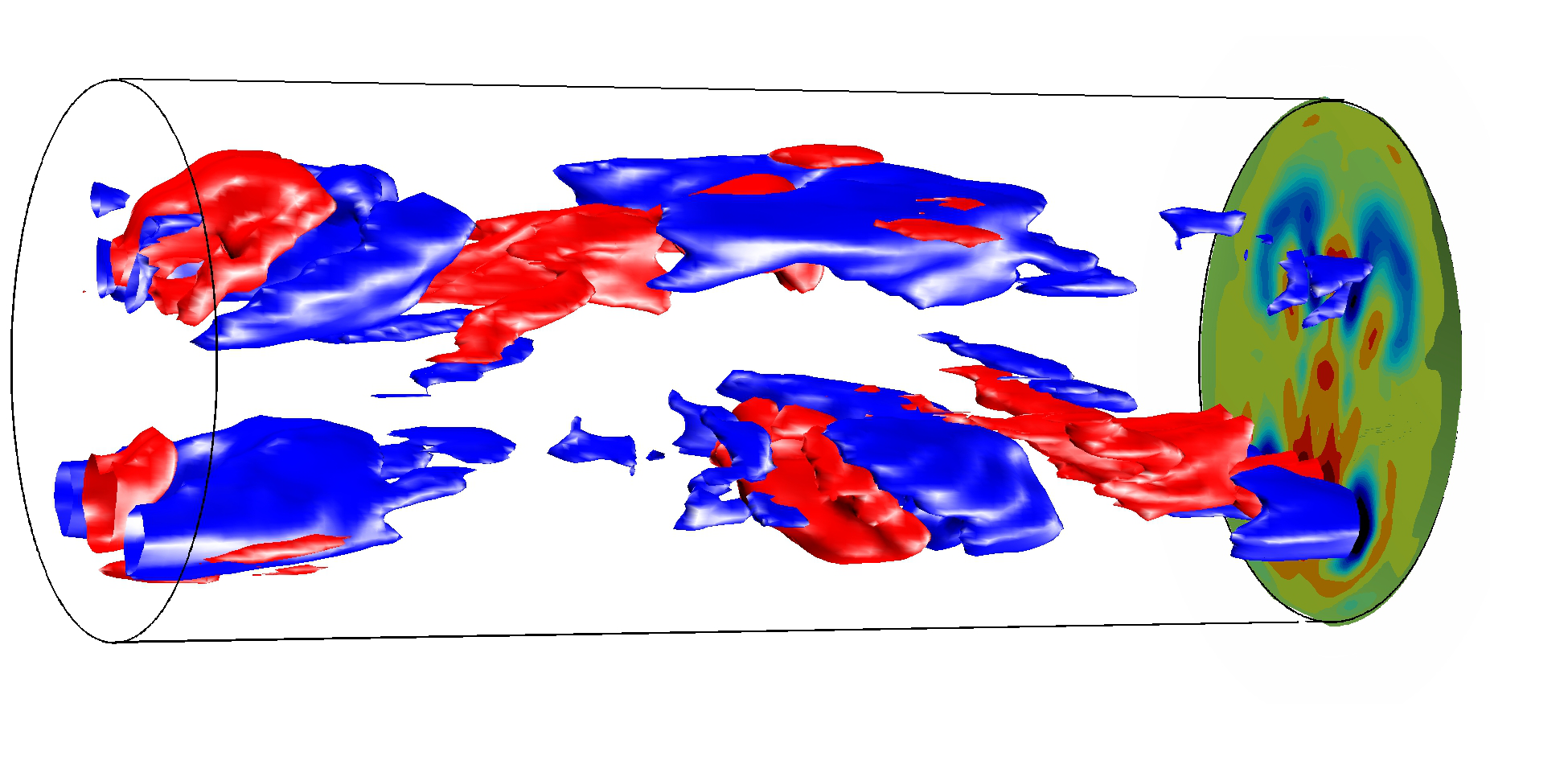}  
  \end{tabular}
  \caption{
           Evolution cycle of the spatio-temporal structures
           of RPO$_{72.001}$ ($T=72.001$) and a turbulent flow hanging
           around it (for about three periods of RPO$_{72.001}$) in the bubble.  The
           phases for the visualisation are identified by the symbols
           $(a)$,$(b)$,$(c)$ in the phase plane $(E,I)$ $(1)$ and
           the monitored energy $E_{3d}$ $(2)$ (see also Figs.  \ref{fig:D-E-I}
           and \ref{fig:E-t_RPOs}).  Flow structures are visualised over one
           full cycle at three times with a constant interval about 24
           $t[D/U]$.  Vortex structures are presented by isosurfaces of
           streamwise vorticity $\omega_z$ at $\pm 0.6\,\text{max}(\omega_z)$
           (red (light gray) is positive and blue (dark gray) is negative) relative to the laminar
           flow for RPO$_{72.001}$ ({\em left column}) and nearby turbulent
           flow ({\em right column}) (From top to bottom they present
           snapshots of $(a)$,$(b)$ and the $(c)$).  Color maps at right
           of each plot present corresponding cross sections (light (darker)
           indicate positive (negative) relative to the laminar flow.). 
           See also online available material movie1.avi.
          } 
\label{fig:comp_RPO_turb}
\end{figure}

Figure \ref{fig:comp_RPO_turb} depicts a full cycle of the temporal
evolution of spatial structures for RPO$_{72.001}$ and a turbulent flow in
its neighbourhood in dark region in Fig.\ \ref{fig:D-E-I}.  
The vortex structures given by isosurfaces of
streamwise vorticity $\omega_z$ 
(constant value for all snapshots of 60$\%$ of its maximum)
show significant variation over one period of RPO$_{72.001}$.  While at the
low energy point of the cycle (see point $(a)$ in phase space plots) the
shape is quite smooth with moderate vorticity. It significantly increases
during the cycle at higher energy levels $(b)$ (see also online available
material), then structure breaks up $(c)$.
There is clear
correspondence between the vortex structures of RPO$_{72.001}$ and the
turbulent flows nearby.  For both cases one observes the same formation
sequence that changes from a streamwise 
vortex-dominant shape to the form of
low-velocity streaks dominating the structure (see movie1.avi).  
This supports the notion that RPOs can encompass much of
the structure and dynamics of turbulent flows.

%%%%%%%%%%%%%%%%%%%%%%%%%%%%%%%%%%%%%%%%%%%%%%%%%%%%%%%%%%%%%%%%%%%%%%
\section{Conclusion}\label{SEC:conclusion}
%%%%%%%%%%%%%%%%%%%%%%%%%%%%%%%%%%%%%%%%%%%%%%%%%%%%%%%%%%%%%%%%%%%%%%

In summary, applying a minimal set of symmetry
we have discovered exact coherent solutions of the
Navier-Stokes equations in pipe flow
and extracted RPOs directly from turbulent velocity fields.  
We have discovered an RPO with long-period period of $72.001\,D/U$, much longer than
previously discovered in pipe flow and comparable shear flows.  
The `bubble' around the orbit is a chaotic saddle within the wider
saddle of turbulence, and
supports the idea that bubbles of chaos continue to have 
and influence {\em after} the crisis bifurcation \citep{KeEc2012}.  
Here, the stable manifold of inner and outer saddles are observed to be highly
intertwined, resulting in entry and exit to the bubble.  
At the initially
chosen parameters we were able to observe the switching process clearly, 
while from lower to higher $\Rey$ we have calculated that there is an
increase in the mean lifetime.
Trajectories that visit the bubble have
significantly longer characteristic lifetimes than those that do not appear to
visit this region, and we have seen that the lifetime is dependent 
on the rate of switching (i.e.\ frequency of visits) to the bubble, rather 
than simply on the lifetime within the bubble itself.
As a general picture, the state space is expected to be filled with a multitude 
of unstable invariant solutions, and turbulent trajectories visit the
least unstable solutions.
The long-period RPO and observations of shadowing
manifests the perception that RPOs capture much of the natural measure of
turbulent flows, within the subregion at least.  Such RPOs are the 
most likely to be extracted directly from simulation.
In future, the natural place to extend these ideas is to the larger domain,
where localised structures such as puffs may exist.  
Work by Avila at al.\ (private communication \citep{Cargese2014}) 
for lower $\Rey$ but larger domains, appear to also be supportive.
It is expected that 
the identification of RPOs will be a valuable tool in decoding
natural characteristics of turbulence.

%%%%%%%%%%%%%%%%%%%%%%%%%%%%%%%%%%%%%%%%%%%%%%%%%%%%%%%%%%%%%%%%%%%%%%
\section*{Acknowledgments}
The research leading to these results has received funding from
the Deutsche Forschungsgemeindschaft (Project No. FOR1182), and the
European Research Council under European Union's Seventh Framework
Programme (FP/2007-2013)/ERC Grant Agreement 306589.
APW is supported by the EPSRC, grant EP/K03636X/1.

%%%%%%%%%%%%%%%%%%%%%%%%%%%%%%%%%%%%%%%%%%%%%%%%%%%%%%%%%%%%%%%%%%%%%%

\bibliographystyle{jfm}

\bibliography{jfm-bibfile}

\end{document}